\documentclass[3p,times,procedia]{elsarticle}
\usepackage{nupha_ecrc}
\usepackage{amsmath,amssymb}

\volume{00}

\firstpage{1}

\journalname{Nuclear Physics A}

\runauth{}

\jid{nupha}

\jnltitlelogo{Nuclear Physics A}

\usepackage{amssymb,bm}

\usepackage[figuresright]{rotating}

\usepackage{slashed}
\usepackage{mathrsfs}

\renewcommand\a{\alpha}
\renewcommand\b{\beta}
\renewcommand\d{\delta}

\renewcommand\l{\lambda}
\renewcommand\r{\rho}

\renewcommand\t{\tau}

\renewcommand\j{\psi}
\renewcommand\o{\omega}
\newcommand\e{\epsilon}
\newcommand\g{\gamma}
\newcommand\z{\zeta}
\newcommand\m{\mu}
\newcommand\n{\nu}
\newcommand\x{\xi}
\newcommand\p{\pi}
\newcommand\h{\theta}
\newcommand\s{\sigma}
\newcommand\f{\phi}
\newcommand\w{\eta}

\renewcommand\L{\Lambda}

\renewcommand\S{\Sigma}
\renewcommand\O{\Omega}

\newcommand\D{\Delta}
\newcommand\G{\Gamma}

\newcommand{\fig}[1]{Fig.~\ref{#1}}
\newcommand{\eq}[1]{Eq.~(\ref{#1})}
\newcommand{\sect}[1]{Sec.~\ref{#1}}
\newcommand{\eqs}[2]{Eqs.~(\ref{#1})-(\ref{#2})}

\newcommand\lb{\left(}
\newcommand\rb{\right)}
\newcommand\ls{\left[}
\newcommand\rs{\right]}

\newcommand{\lan}{\langle}
\newcommand{\ran}{\rangle}

\newcommand\ra{\rightarrow}

\newcommand{\non}{\nonumber\\}
\newcommand\pt{\partial}

\newcommand{\Tr}{{\rm Tr}}

\newcommand{\bp}{{\vec p}}

\newcommand{\bv}{{\vec v}}

\renewcommand{\part}{{\rm part}}

\usepackage{slashed}
\usepackage{mathrsfs}

\newcommand{\tr}{ {\rm Tr} }

\newcommand\mc{\mathcal}

\newcommand\na{\nabla}
\newcommand\ola{\overleftarrow}

\renewcommand{\vec}{\boldsymbol}
\newcommand\be{\begin{equation}}
\newcommand\ba{\begin{eqnarray}}
\newcommand\ee{\end{equation}}
\newcommand\ea{\end{eqnarray}}

\begin{document}

\begin{frontmatter}



\dochead{XXVIIIth International Conference on Ultrarelativistic Nucleus-Nucleus Collisions\\ (Quark Matter 2019)}

\title{Vorticity and Spin Polarization --- A Theoretical Perspective}

\author[1,2]{Xu-Guang Huang}
\address[1]{Physics Department and Center for Particle Physics and Field Theory, Fudan University, Shanghai 200433, China}
\address[2]{Key Laboratory of Nuclear Physics and Ion-beam Application (MOE), Fudan University, Shanghai 200433, China}
\begin{abstract}
  We give a theoretical perspective on the vorticity and spin polarization in heavy-ion collisions. We discuss the recent progress in spin hydrodynamics and spin kinetic theory. We also discuss other effects caused by vorticity including the chiral vortical effect and rotation-induced phase transitions.
\end{abstract}

\end{frontmatter}


\section{Introduction}
\label{sec:intro}
Vortices are common phenomena in both classical and quantum fluids. Examples exist in systems across a wide range of scales from the rotating galaxies to the tornadoes on earth to the quantum vortices in superfluids. The strength of a vortex is measured by the vorticity which, in non-relativistic hydrodynamics, is defined as
\begin{eqnarray}
\label{non-vor}
	\bm{\omega}=\frac{1}{2}\bm{\nabla}\times\bm{v},
\end{eqnarray}
where $\bm{v}$ is the flow velocity. In heavy-ion collisions, fluid vortices can be induced by different sources, e.g., the global angular momentum (AM) of the colliding system, the magnetic field, the propagating jets in the quark-gluon matter, and the inhomogeneous expansion of the fireball. 
It is not difficult to imagine that in a non-central collision the system possesses a big AM, which can be estimated as $J\sim\sqrt{s}Ab/2\sim 10^{6}\hbar$ for Au + Au collision at $\sqrt{s}=200$ GeV at impact parameter $b=10$ fm where $A$ is the nucleon number of the ion. After the collision, a portion of this AM is retained by the produced quark-gluon matter in the form of fluid vorticity. In 2017, the STAR Collaboration published the first evidence of the vorticity via the measurement of the spin polarization of $\Lambda$ and $\bar{\Lambda}$ hyperons (``$\Lambda$ polarization" hereafter) in Au + Au collisions~\cite{STAR:2017ckg}. The experimental result can be well described by the theoretical calculations based on the vorticity interpretation of the $\Lambda$ polarization. In 2018 and 2019, new experimental results were reported by STAR Collaboration~\cite{Adam:2018ivw,Adam:2019srw}. These new results contain differential information of the $\Lambda$ polarization which, however, cannot be satisfactorily explained using the vorticity interpretation. 

The strong vorticity may induce other interesting effects. A famous one is the chiral vortical effects (CVEs), i.e., the generation of vector and axial currents along the vorticity. Since the vorticity characterizes the local angular velocity of the fluid cell, the strong vorticity found in heavy-ion collisions also inspires the study of quantum chromodynamics (QCD) phase transitions under rotation. We will also give a brief overview of the CVEs, the chiral vortical wave (CVW) associated with CVEs, and the rotation-induced phase transitions.

\section{Vorticity in heavy-ion collisions}
\label{sec:vort}
For relativistic fluid, like the quark-gluon plasma produced in heavy-ion collisions, different vorticities can be introduced for different physical applications. We discuss two examples here. (1) The kinematic vorticity, which is a covariant generalization of the definition (\ref{non-vor}):
\begin{eqnarray}
\label{def:kin}
\o^\m=\frac{1}{2}\e^{\m\n\r\s}u_\n\pt_\r u_\s,
\end{eqnarray}
where $u^\m=\g(1,\bm v)$ is the flow four velocity with $\g=1/\sqrt{1-\bv^2}$ the Lorentz factor. Its tensorial form is $\o_{\m\n}=(1/2)(\pt_\n u_\m-\pt_\m u_\n)$ which links to $\o^\m$ by $\o^\m=-(1/2)\e^{\m\n\r\s}u_\n\o_{\r\s}$. (2) The thermal vorticity:
\begin{eqnarray}
\varpi_{\m\n}=\frac{1}{2}[\pt_\n(\b u_\m)-\pt_\m(\b u_\n)],
\end{eqnarray}
where $\b=1/T$ is the inverse temperature. The importance of thermal vorticity is that it characterizes the global equilibrium of a rotating fluid and determines the strength of the spin polarization at global equilibrium~\cite{Becattini:2013fla,Fang:2016vpj,Liu:2020flb}. We will show this in \sect{sec:kine}.

\begin{figure}[h]
\begin{center}
\includegraphics[height=4.0cm]{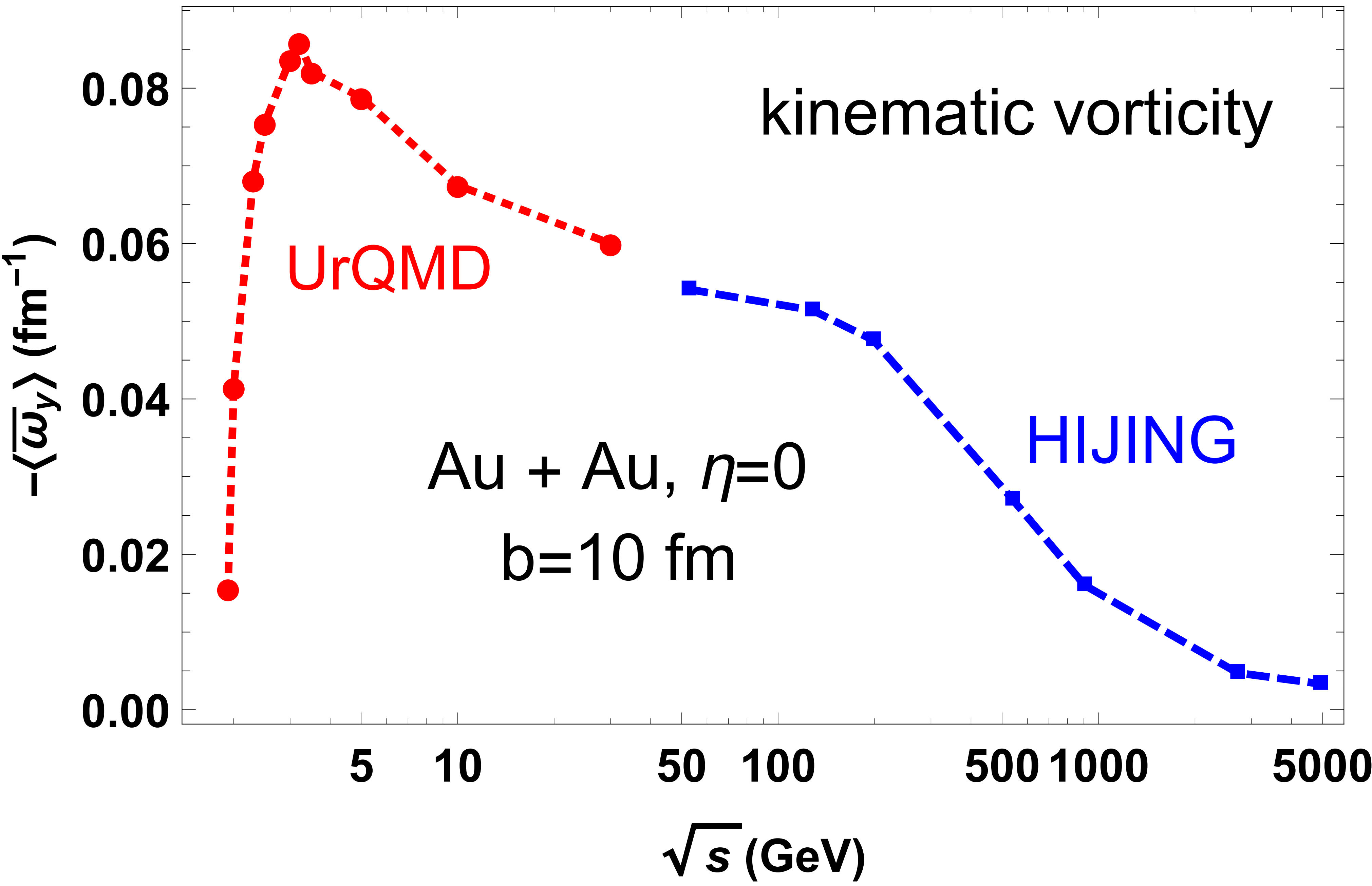}\;\;\;\;\;\;\;\;\;\;\;\;\;
\includegraphics[height=4.0cm]{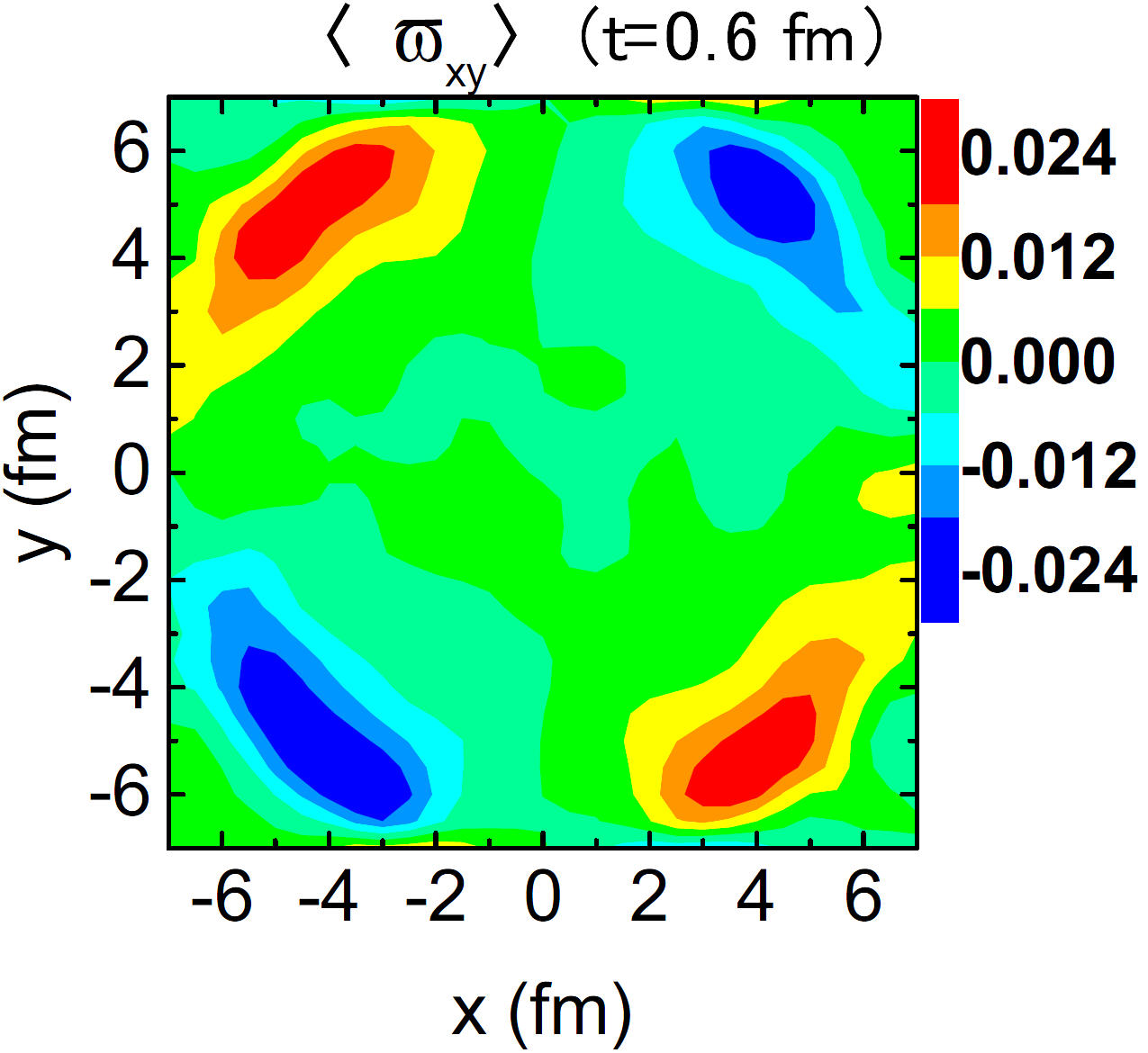}
\caption{(Left) Kinematic vorticity (averaged over transverse overlapping region at mid-rapidity and over events) versus $\sqrt{s}$~\cite{Deng:2016gyh,Deng:2020ygd}. (Right) The distribution of thermal vorticity $\lan\varpi_{xy}\ran$ in transverse plane ($z=0$) for Au + Au collisions at $\sqrt{s}=19.6$ GeV averaged over the centrality region 20-50\%~\cite{Wei:2018zfb}.}
\label{vor-eng}
\end{center}
\end{figure}
\begin{figure}[h]
\begin{center}
\includegraphics[height=3.5cm]{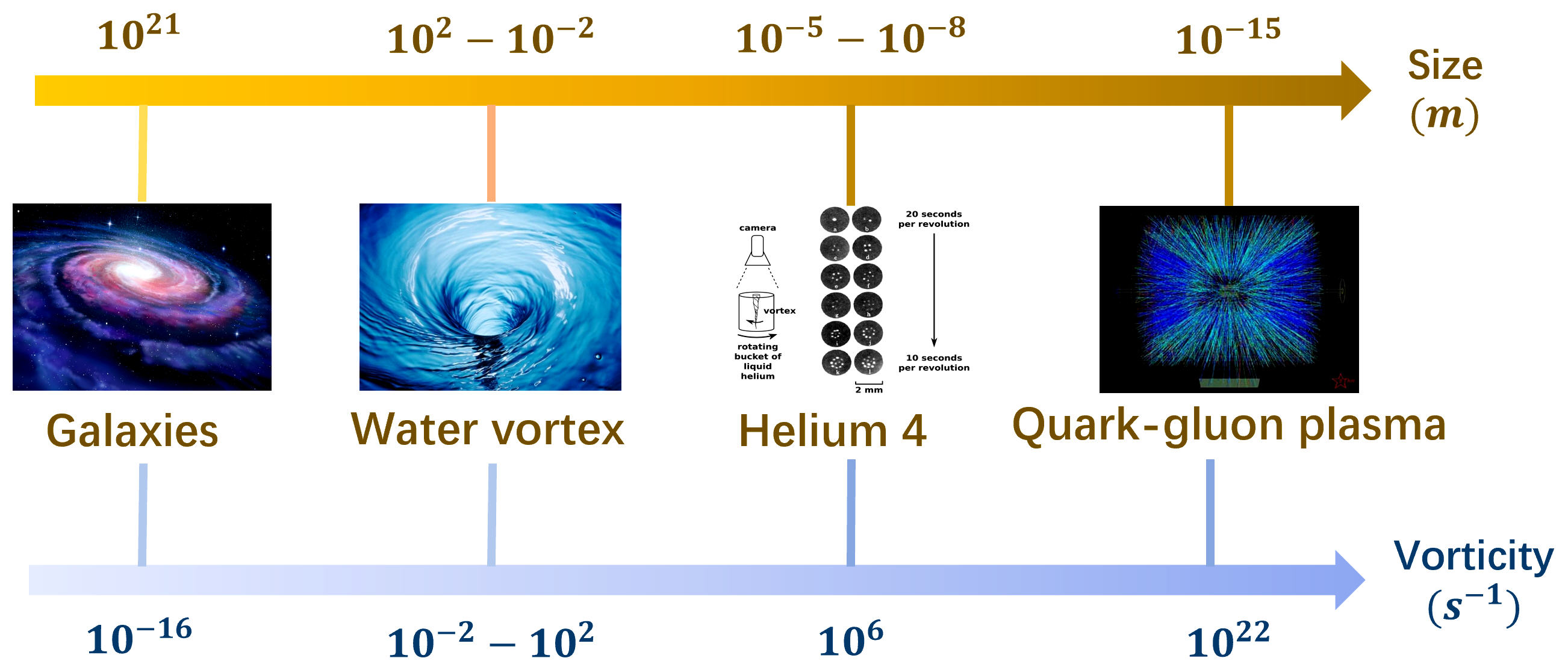}
\caption{A carton for comparison of the vorticity versus system size in nature.}
\label{carton}
\end{center}
\end{figure}

In \fig{vor-eng} (Left), we show the collision energy dependence of the $y$-component of the kinematic vorticity for Au + Au collisions at fixed impact parameter $b=10$ and rapidity $\eta=0$, where $\lan\bar\o_y\ran$ means average over transverse overlapping region weighted by energy density and over events. The low-energy data are obtained by using UrQMD model (red dotted line) for initial time (defined as the moment when the longitudinal density maximizes)~\cite{Deng:2020ygd} and high-energy data are obtained by using HIJING model (blue dashed line) for $\t=0.4$ fm~\cite{Deng:2016gyh}. Similar calculations can also be found in Refs.~\cite{Jiang:2016woz,Ivanov:2017dff}. The simulations show:
\vspace{-0.25cm}
\begin{itemize}
\item The vorticity is very strong with typical strength $\sim 10^{21}-10^{22}$ s$^{-1}$ consistent with the experimental extraction~\cite{STAR:2017ckg}. In this sense, the heavy-ion collisions create ``the most vortical fluid"; see \fig{carton}.
\vspace{-0.25cm}
\item The initial $\lan\bar\o_y\ran$ at $\w=0$ first increases with $\sqrt{s}\gtrsim 2m_N$ ($m_N$: nucleon mass) because the AM at $\w=0$ increases, then decreases at higher energy because the matter at $\w=0$ becomes more Bjorken boost invariant and supports less vorticity --- a feature also shown in data of global $\L$ polarization (i.e., the mean spin polarization over all $\L$ or $\bar\L$ at mid-rapidity) .
\end{itemize}

In \fig{vor-eng} (Right), we show the numerical simulation of the longitudinal component of the event-averaged thermal vorticity in the transverse plane at mid-rapidity~\cite{Wei:2018zfb}; see also Refs.~\cite{Becattini:2017gcx,Xie:2019npz}. The special quadrupolar pattern of $\lan\varpi_{xy}\ran$ emerges due to the inhomogeneous and anisotropic transverse expansion of the fireball (superposed by the finite transverse gradient of temperature). It may be related to the analogous quadrupolar distribution of longitudinal $\L$ polarization which we will discuss in next section.

\section{Spin polarization}
\label{sec:spin}
\begin{figure}[h]
\begin{center}
\includegraphics[height=4.0cm]{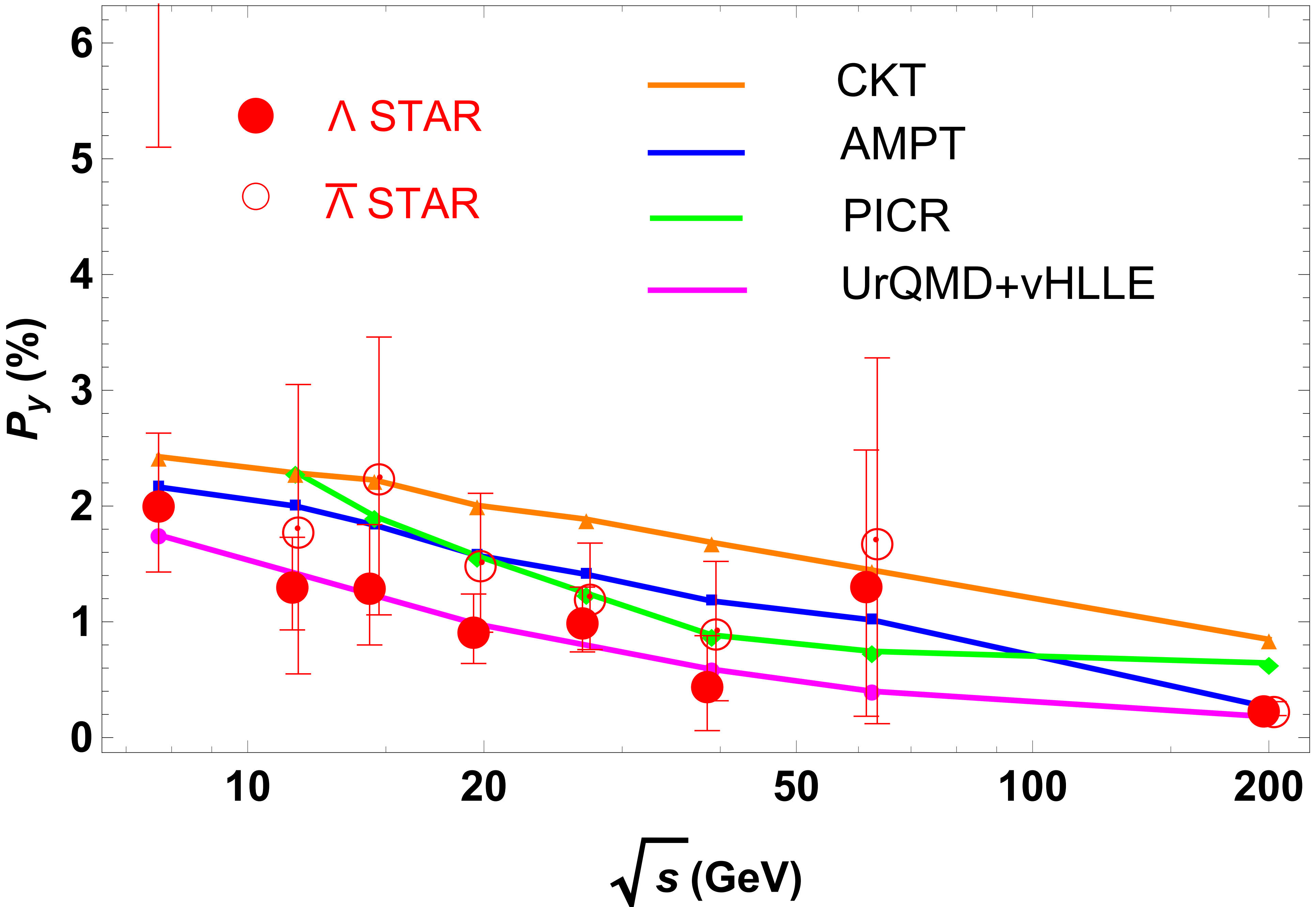}
\caption{The global $\L$ polarization: Experimental data~\cite{STAR:2017ckg,Adam:2018ivw} versus results from four different numerical approaches~\cite{Wei:2018zfb,Sun:2017xhx,Xie:2017upb,Karpenko:2016jyx}.}
\label{globpolar}
\end{center}
\end{figure}

A remarkable effect of the vorticity is that it could polarize the spin of the constituent particles through the quantum mechanical spin-orbit coupling~\cite{Liang:2004ph,Becattini:2007sr,Gao:2007bc,Huang:2011ru}. This can be easily seen by considering a thermal equilibrium state under rotation (the vorticity is just the rotating frequency in this case). The density operator is $\hat{\r}=Z^{-1}\exp\ls-\b( \hat{H}- \hat{\bm S}\cdot\bm \o)\rs$ with $\hat{H}, Z$, and $\hat{\bm S}$ the spin-unpolarized Hamiltonian, partition function, and spin operator. The spin polarization, given by $\bm P=\Tr[\hat{\bm S}\hat{\r}]/s$ ($s\neq0$: the spin quantum number), thus reads $\bm P=(s+1)\bm\o/(3T)+O[(\o/T)^2]$. This heuristic argument can be made rigorous and the polarization four-vector in phase space is (for spin-1/2 fermions)~\cite{Becattini:2013fla,Fang:2016vpj,Liu:2020flb}
\begin{eqnarray}
\label{spin}
P^\m(x,p)=-\frac{1}{4m}(1-n_F)\e^{\m\n\r\s}p_\n\varpi_{\r\s}(x)+O(\varpi^2),
\end{eqnarray}
where $n_F=n_F(x,p)$ is the Fermi-Dirac distribution function and $\varpi_{\r\s}(x)$ is the thermal vorticity. Integrating over a freeze-out hypersurface $\S_\r$, we obtain the spin polarization of particles with four-momentum $p$:
\begin{eqnarray}
\label{spin2}
P^\m(p)=-\frac{1}{4m}\e^{\m\n\r\s}p_\n\frac{\int d\S_\r p^\r n_F(1-n_F)\varpi_{\r\s}(x)}{\int d\S_\r p^\r n_F}+O(\varpi^2).
\end{eqnarray}
In experiments, the spin polarization is measured in the rest frame of the particle so that $P^{\m}=(0,\bm P)$ which links to the spin polarization in the laboratory frame by a Lorentz transformation,
$\bm P\ra\bm P-(\bm p\cdot\bm P)\bm p/[E(E+m)]$,
where $E=\sqrt{\bp^2+m^2}$. Equations~(\ref{spin})-(\ref{spin2}) are often used in numerical calculations; in particular, \eq{spin} [\eq{spin2}] is suitable for transport (hydrodynamic) model simulations. In \fig{globpolar}, we show the results of global $\L$ polarization from different theoretical approaches including chiral kinetic theory (CKT)~\cite{Sun:2017xhx}, AMPT transport model~\cite{Wei:2018zfb}, PICR hydrodynamic model~\cite{Xie:2017upb}, and UrQMD+vHLLE hybrid model~\cite{Karpenko:2016jyx}; see also Refs.~\cite{Li:2017slc,Shi:2017wpk,Ivanov:2019ern}. All the numerical results are for the polarization of primary $\L$, and $\bar{\L}$; the feed-down effect can give a $\sim 10\%$ suppression~\cite{Xia:2019fjf}. The theoretical results fit well the experimental data, strongly supporting the vorticity interpretation of the global spin polarization.

One can also use \eqs{spin}{spin2} to calculate the differential spin polarization, namely, the dependence of $\L$ polarization on kinematic variables, particularly, the azimuthal angle $\phi$, which can be expressed by a harmonic expansion:
\begin{eqnarray}
\label{spin-har}
\frac{d\bm P}{d\phi}=\bm P+2 \bm f_{2}\sin[2(\phi-\Psi_{\rm RP})]
+2 \bm g_{2}\cos[2(\phi-\Psi_{\rm RP})]+\cdots,
\end{eqnarray}
where $\Psi_{\rm RP}$ is the reaction-plane angle. The second-order harmonic coefficients $\bm f_{2}$ and $\bm g_{2}$ contain information of the local spin polarization. For example, the local vorticity distribution shown in \fig{vor-eng} (Right) leads to similar $\L$-polarization distribution in momentum space via \eq{spin2} and gives rise to $f^{\rm ther}_{2z}<0$~\cite{Wei:2018zfb,Becattini:2017gcx,Xia:2018tes}. Similarly, for the $y$-component of $\L$ polarization, theoretical calculations show that $g^{\rm ther}_{2y}<0$~\cite{Wei:2018zfb,Xie:2016fjj}. Recently, the STAR collaboration published the measurement of $dP_{y,z}/d\f$~\cite{Adam:2018ivw,Adam:2019srw} which gives $f_{2z}^{\rm exp}>0, g_{2y}^{\rm exp}>0$, opposite to the theoretical calculations. This raises a spin ``sign problem" which challenges the primitive equilibrium vorticity interpretation of the spin polarization. In order to resolve the spin sign problem, several important ingredients should be carefully (re)-examined:
\vspace{-0.2cm}
\begin{itemize}
\item About $80\%$ of the measured $\L$ and $\bar{\L}$ are from decays of higher-lying hadrons. Some decay channels (e.g., $\S^0\ra\L+\g$) can even flip the spin-polarization direction of the daughter $\L$ comparing to the parent particle. The recent studies showed that such decay contributions, though suppress $\sim 10\%$ of the primary $\L$ polarization, are not enough to resolve the spin sign problem~\cite{Xia:2019fjf,Becattini:2019ntv}.
\vspace{-0.23cm}
\item The possible initial local spin polarization or initial flow profile that can lead to finite local vorticity have not been encoded in currently established hydrodynamic or transport models. It is a desirable task to perform a numerical test of such possible initial conditions.
\vspace{-0.23cm}
\item The formulas (\ref{spin}) and (\ref{spin2}) are derived under the assumption that both momentum and spin degree of freedom reach global equilibrium which, however, may not be the realistic case in heavy-ion collisions. Away from global equilibrium, spin polarization is no longer enslaved to thermal vorticity and should be treated as an independent dynamical variable. We will discuss the hydrodynamic and kinetic frameworks with spin as dynamical variable in \sect{sec:hydr} and \sect{sec:kine}.
\vspace{-0.23cm}
\item Other ingredients that may influence the $\L$ polarization should also be explored, e.g., the magnetic fields~\cite{Guo:2019joy}, the hadronic mean-fields~\cite{Csernai:2018yok}, the chiral-anomaly induced effects~\cite{Sun:2018bjl,Liu:2019krs}, the other possible spin chemical potentials~\cite{Florkowski:2019voj,Wu:2019eyi}, and the gluonic contribution. It is also helpful to examine complementary observables for measuring the vorticity, e.g., the $\phi$- and $K^{*0}$-spin alignment~\cite{Liang:2004xn}, the CVEs and CVW, and the recently-proposed vorticity-dependent hadron yields~\cite{Taya:2020}.
\end{itemize}

\section{Spin hydrodynamics}
\label{sec:hydr}
Spin hydrodynamics and spin kinetic theory (SKT) are systematic frameworks to describe the spin polarization away from global equilibrium. They are under rapid development. We summarize the recent progress here.

In spin hydrodynamics, the spin polarization density (or equivalently the spin chemical potential $\O^{\m\n}$; see below) is treated as a (quasi-)hydrodynamic variable, on similar footing as the temperature $T$ and flow velocity $u^\m$~\cite{Florkowski:2017ruc,Montenegro:2018bcf,Florkowski:2018fap,Hattori:2019lfp,Bhadury:2020puc}.
In the so-called first order theory, the energy-momentum tensor and the spin current tensor are given by (in Landau-Lifshitz frame),
\begin{eqnarray}
\label{spin-cons}
&T^{\m\n}=eu^\m u^\n-P\D^{\m\n}+\s_\w^{\m\n}+\s_\z^{\m\n}+2q^{[\m}u^{\n]}+\f^{\m\n},\nonumber\\
&\S^{\m,\a\b}=u^\m S^{\a\b},
\end{eqnarray}
where $e$ is the energy density, $P$ is the pressure, $\s^{\m\n}_\w,\s^{\m\n}_\z$  are shear and bulk viscous tensors, $q^{\m}$ and $\f^{\m\n}=\f^{[\m\n]}$ with $X^{[\a\b]}=(X^{\a\b}-X^{\b\a})/2$ are spin-related quantities describing the strength of the torque on the temporal and spacial components of the spin current tensor. The constitutive relations read~\cite{Hattori:2019lfp}
\begin{eqnarray}
\label{spin-deco}
&\s_\w^{\m\n}=2\w\pt_\perp^{\lan\m}u^{\n\ran},\\
&\s_\z^{\m\n}=\z\h\D^{\m\n},\\
&q^\m=\l(D u^\m+\b\pt_\perp^\m T-4\O^{\m\n}u_\n),\\
\label{spin-deco2}
&\f^{\m\n}=2\g(\pt_\perp^{[\m}u^{\n]}+2\O^{\m\n}_\perp),
\end{eqnarray}
where $X^{\lan\a\b\ran}=(X^{\a\b}+X^{\b\a})/2-{X^\m}_\m\D^{\a\b}/3$, $\h=\pt_\m u^\m$ is the expansion rate, $\D_{\m\n}=g_{\m\n}-u_\m u_\n$ is the spatial projection, $D=u\cdot\pt$ is the co-moving time derivative, $\pt^\m_\perp=\D^{\m\n}\pt_\n$ is the spatial derivative, $\O^{\m\n}$ is called the spin chemical potential, $\O^{\m\n}_\perp=\D_{\m\r}\D_{\n\s}\O^{\r\s}$. Here, $\w, \z, \l, \g$ are the transport coefficients called shear viscosity, bulk viscosity, boost heat conductivity, and rotational viscosity~\cite{Hattori:2019lfp} which must be semi-positive as required by the second law of thermodynamics. The hydrodynamic equations are the energy-momentum and AM conservation laws,
\begin{eqnarray}
\label{spin-hydro}
&\pt_\m T^{\m\n}=0,\\
&\pt_\m \S^{\m,\a\b}=4q^{[\b}u^{\a]}2+\f^{\b\a},
\end{eqnarray}
plus the equation of state which links $e,P,S^{\a\b}$. Note that $S^{\a\b}$ links to $\O^{\a\b}$ intrinsically by definition.

In practical use, the above first-order theory has severe problem due to the appearance of uncausal modes and numerical instability. The easiest way to overcome this problem is to amend \eqs{spin-deco}{spin-deco2} to the Israel-Stewart form,
\begin{eqnarray}
\label{spin-IS}
&\t_\w (D\s_\w^{\m\n})_\perp+\s_\w^{\m\n}=2\w\pt_\perp^{\lan\m}u^{\n\ran},\\
&\t_\z (D\s_\z^{\m\n})_\perp+\s_\z^{\m\n}=\z\h\D^{\m\n},\\
&\t_\l (Dq^\m)_\perp+q^\m=\l(D u^\m+\b\pt_\perp^\m T-4\O^{\m\n}u_\n),\\
&\t_\g (D\f^{\m\n})_\perp+\f^{\m\n}=2\g(\pt_\perp^{[\m}u^{\n]}+2\O^{\m\n}_\perp),
\label{iseq}
\end{eqnarray}
where $(\cdots)_\perp$ means the components transverse to $u^\m$, e.g., $(D\s_\w^{\m\n})_\perp=\D^{\m}_\r\D^\n_\s D\s_\w^{\r\s}$ and $\t_\w, \t_\z, \t_\l, \t_\g$ are relaxation times. With given initial conditions for $T, u^\m, \O^{\a\b}, \s_\w^{\m\n}, \s_\z^{\m\n}, q^\m, \f^{\m\n}$, \eqs{spin-hydro}{iseq} and the equation of state form a set of closed, numerically stable, differential equations. It would be an important future task to numerically apply the above spin hydrodynamics to heavy-ion collisions which would provide valuable insights into the spin sign problem. Besides, some theoretical issues need deeper study, e.g., the pseudo-gauge ambiguity of defining the spin current tensor~\cite{Becattini:2018duy}, the development of the full second-order theory, the situation with $O[(\pt)^0]$-order vorticity, and the calculation of new transport coefficients.

\section{Spin kinetic theory}
\label{sec:kine}
In addition to hydrodynamics, kinetic theory is another widely used framework to study many-body system out of equilibrium. Consider Dirac fermions whose Wigner function is (in this section, we consider a background curved spacetime and electromagnetic field and restore $\hbar$)
\ba
W(x,p)=\int\sqrt{-g(x)}d^4y e^{-ip\cdot y/\hbar} \bar{\j}\lb x,\frac{y}{2}\rb \otimes \j\lb x,-\frac{y}{2}\rb,
\label{wp}
\ea
where $\j(x,y) \equiv e^{y \cdot D}\j(x)$ with $\j(x)$ the spinor and $D_\m$ the covariant derivative in tangent bundle and $U(1)$ bundle [e.g., $D_\m\j(x,y)=(\na_\m-\G^{\l}_{\m\n}y^\n \pt_\l^y+iA_\m/\hbar)\j(x,y)$]. Applying the Dirac equation $ [i\hbar\gamma^\mu(\nabla_\mu + iA_\mu/\hbar)-m]\,\psi(x) = \bar{\psi}(x)\,[i\hbar({\ola\nabla}_\mu - iA_\mu/\hbar)\gamma^\mu+m] = 0$ and making a $\hbar$ expansion, one can find that the full dynamics of $W(x,p)$ is controlled by the vector and axial currents $\mc{V}^\m=\tr\lb\g^\m W\rb$ and $\mc{A}^\m=\tr\lb\g^\m\g_5 W\rb$. (1) For massive case, they take the forms
\ba
&\mc{V}^\m
= 4\p \bigg\{ p^\m f \d(p^2-m^2)
+m\hbar \tilde{F}^{\m\n} \h_\n f_A \d'(p^2-m^2)+\frac{\hbar}{2m}\epsilon^{\m\n\r\s} p_\n \D_\r \lb \h_\s f_A\rb \d(p^2-m^2)\bigg\},
\label{solutionmsv}
\\
&\mc{A}^\m = 4\p\big\{m\h^\m f_A \d(p^2-m^2)+ \hbar \tilde{F}^{\m\n}p_\n f \d'(p^2-m^2)\big\},
\label{mssa}
\ea
where $f=f_++f_-$, $f_A=f_+-f_-$, and $\Delta_\mu = \nabla_\mu +(-F_{\mu\lambda}+\G^\n_{\m\l}p_\n)\partial_p^\lambda$. The physical meanings of $\h^\m$ and $f_\pm$ are the following: $\h^\m$ is the spin quantization direction satisfying $\h^2=-1$ and $p\cdot\h=0$, $f_\pm$ is the distribution function of spin along/anti-along $\h^\m$.
The kinetic equations at $O(\hbar)$ are as follows~\cite{Liu:2020flb}:
\ba
\label{keosspud}
&\!\!\!\!\!\!\!\d(p^2-m^2\mp \hbar\S_S^{\a\b} F_{\a\b})\bigg\{\bigg[p^\m \D_\m
\pm\frac{\hbar}{2}\S_S^{\m\n} \lb \na_\r F_{\m\n} -p_\l{R^\l}_{\r\m\n} \rb \pt^\r_p\bigg] f_{\pm}+\frac{\hbar}{2}f_A\lb \na_\r F_{\m\n} -p_\l{R^\l}_{\r\m\n}\rb \pt^\r_p\S_S^{\m\n} \bigg\}=0,\\
&\d(p^2-m^2)\bigg[f_Ap\cdot\D \h^\m -
f_AF^{\m\n} \h_\n
+ \h^\m  p\cdot\D f_A -\frac{\hbar}{4m }\epsilon^{\m\n\r\a} p_\a \lb \na_\s F_{\n\r} -p_\l{R^\l}_{\s\n\r} \rb \pt^\s_p f\bigg]=0,
\label{keospin}
\ea
where $\S_S^{\m\n}=\frac{1}{2m}\e^{\m\n\r\s}\h_\r p_\s$ is the spin tensor for massive fermions. More discussions in Minkowski spacetime can be found in Refs.~\cite{Gao:2019znl,Weickgenannt:2019dks,Hattori:2019ahi,Wang:2019moi,Yang:2020hri}. (2) For massless case, up to $O(\hbar)$, $\mc{V}^\m$ and $\mc{A}^\m$ read
\ba
\lb \mc{V},\mc{A}\rb^\m=4\p \big\{ \ls p^\m \lb f,f_5 \rb
+\hbar \S_n^{\m\n}\D_\n \lb f_5,f \rb \rs  \d(p^2)+ \hbar \tilde{F}^{\m\n}p_\n \lb f_5,f \rb \d'(p^2)\big\},
\label{masslessaxial}
\ea
where $f=f_R+f_L$ and $f_5=f_R-f_L$ with $f_{R/L}$ the right-/left-hand distribution function and $\S_n^{\m\n}=\frac{1}{2p\cdot n}\e^{\m\n\r\s}p_\r n_\s$ is the spin tensor for massless fermions with $n^\m$ a unit timelike frame vector. The evolution of $f_{R/L}$ is controlled by the chiral kinetic equation~\cite{Liu:2018xip}:
\ba
\d(p^2\mp\hbar F_{\a\b}\S_n^{\a\b}) \bigg[p_\m \D^\m  f_{R/ L}\pm\frac{\hbar}{p\cdot n}\tilde{F}_{\m\n} n^\m \D^\n f_{R/ L}
\pm \hbar \D^\m \lb \S^n_{\m\n}\D^\n f_{R/ L} \rb\bigg]=0.
\label{keml}
\ea

Let us focus on the massive case. The local equilibrium state is specified by distributions $f^{\rm LE}_{\pm}=n_F(g_{\pm})$ with $g_{\pm}=p \cdot \b + \a_\pm \pm \hbar\S_S^{\m\n}\O_{\m\n}$ ($\b, \a_\pm, \O_{\m\n}$ depend on $x$ only). Furthermore, one can find that the following condition
\ba
&\na_\m \b_\n + \na_\n \b_\m = 0,\;\;\;\;
\na_{[\m} \b_{\n]} - 2\O_{\m\n} = 0,\;\;\;
\na_\m \a_{\pm} =F_{\m\n}\b^\n,\non
&\a_+ = \a_-,\;\;\;\;
\h^\m=-\frac{1}{2m\G}\e^{\m\n\r\s}p_\n\nabla_{[\r}\b_{\s]},\nonumber
\label{eqspinvec}
\ea
where $\G^2=\frac{1}{2}\na_{[\m}\b_{\n]}\L^{\m\r}\L^{\n\s}\na_{[\r}\b_{\s]}$ with $\L^{\m\n}=g^{\m\n}-p^\m p^\n/m^2$, fulfils \eqs{keosspud}{keospin}. The state specified by the above condition is called the global equilibrium state. The spin polarization per particle in phase space is defined by the Pauli-Lubanski vector divided by $s=1/2$ which can be reduced to $P^\m=\mc{A}^\m/(4\p sf)$~\cite{Liu:2020flb}. At global equilibrium,  integrating over energy for \eq{mssa}, we obtain (here we consider Minkowski spacetime and zero electromagnetic field)
\ba
P^\m_{\rm GE}&=&-\frac{\hbar}{4E} (1-n_F)\e^{\m\n\r\s}p_\n\varpi_{\r\s},
\label{sppmsvpp2}
\ea
which gives formula (\ref{spin}) after approximating $\sqrt{\bp^2+m^2}\approx m$. We note that \eq{sppmsvpp2} holds even for massless fermions~\cite{Liu:2020flb}. It would be important to derive the collision terms for the SKT. Recent attempts are Refs.~\cite{Carignano:2019zsh,Li:2019qkf,Yang:2020hri}.

\section{Chiral vortical effect}
\label{sec:chir}
Within the framework of CKT, substituting \eq{masslessaxial} into the definitions of vector and axial currents $J_V(x)=\int_p \tr(W\g^\m), J_A(x)=\int_p\tr(W\g^\m\g_5)$ with $\int_p=\int d^4p/(2\p)^4$ (Minkowski spacetime), at global equilibrium, one finds~\cite{Liu:2018xip,Huang:2018aly}
\ba
J^\m_V=\frac{\m_V\m_A}{\p^2}\o^\m, \;\;\;\; J^\m_A=\lb\frac{\m^2_V+\m^2_A}{2\p^2}+\frac{T^2}{6}\rb\o^\m,
\ea
where $\o^\m$ is the kinematic vorticity and $\m_{V,A}$ is the vector/axial chemical potential. These are the CVE currents~\cite{Erdmenger:2008rm,Banerjee:2008th}. Experimentally, the vector CVE could induce a baryonic current along the vorticity which leads to an event-by-event baryon-anti-baryon separation with respect to the reaction plane. A possible observable is the two-particle correlation $\w_{\a\b}=\lan\cos(\phi_\a+\phi_\b-2\Psi_{\rm RP})\ran$ where $\a,\b=\pm$ denote baryons or anti-baryons and $\phi_{\a,\b}$ are the corresponding azimuthal angles. Although the data from STAR Collaboration show features consistent with the expectation of the vector CVE~\cite{Zhao:2014aja}, the $\w$-correlation contains background contributions from, e.g., local baryon number conservation which is challenging to subtract. Another possible observable of CVEs is that the CVEs can induce two propagating wave modes along the vorticity which are called chiral vortical waves (CVWs). They transport baryonic charge in such a ways that more baryons are distributed on the tips of the fireball and more anti-baryons in the equator of the fireball. This would lead to a larger elliptic flow ($v_2$) for anti-baryons (say, $\bar{\L}$) and baryons (say, $\L$) with the difference proportional to the net baryon asymmetry $A_\pm^{\L}=(N_\L-N_{\bar\L})/(N_\L+N_{\bar\L})$~\cite{Jiang:2015cva}. As $\L$ and $\bar{\L}$ are rare in heavy-ion collisions, the detection of this difference is statistically challenging. We expect that the phase II of the RHIC beam energy scan program can provide the opportunity for the search of CVWs.

\section{Rotation induced phase transition}
\label{sec:rota}
Since the vorticity is a measure of the local rotation, the existence of strong vorticity in heavy-ion collisions also inspired the studies of QCD phases under rotation. The (uniform) rotation has two fundamental effects: On one hand, it introduces a ``chemical potential" for the AM as easily seen from the shift in Hamiltonian, $\hat{H}_{\rm rot}=\hat{H}-\bm\o\cdot\bm J$ ($\bm J$: total AM). On the other hand, a uniformly rotating system must be finite in order to maintain causality~\cite{Ebihara:2016fwa}. This latter effect would induce a finite gap to the fermionic excitations (for certain boundary conditions) making the vacuum inactive to uniform rotation. The combination of the above two effects lead to very interesting consequences: at finite temperature or density or magnetic field (or possibly other external intensive thermodynamic forces), a uniform rotation would suppress the condensate of spin-0 composite quark-quark or quark-anti-quark pairings~\cite{Chen:2015hfc,Jiang:2016wvv,Chernodub:2016kxh,Chernodub:2017ref,Huang:2017pqe,Liu:2017spl,Wang:2018sur,Wang:2018zrn,Wang:2019nhd,Zhang:2018ome,Cao:2019ctl,Chen:2019tcp}. For example, under magnetic field, the rotation tends to suppress the chiral condensate leading to the novel rotational magnetic inhibition~\cite{Chen:2015hfc}.

The study of the rotation-induced phase transition extends the usual QCD phase diagram on temperature-density plane to one additional dimension, the rotation dimension. There are certainly many unexplored issues waiting for investigation, e.g., the influence of rotation on Polyakov loop once the gluonic section is concerned and the induction of magnetization due to strong rotation (relativistic Barnett effect).

{\it Acknowledgments.---} We thank H. L. Chen, W. T. Deng, X. G. Deng, K. Fukushima, L. L. Gao, K. Hattori, M. Hongo, H. Z. Huang, P. Huovinen, H. Li, Y. C. Liu, Y. Jiang, J. Liao, Y. G. Ma, K. Mameda, M. Matsuo, K. Nishimura, L. G. Pang, A. V. Sadofyev, H. Taya, Q. Wang, X. N. Wang, D. X. Wei, H. Z. Wu, N. Yamamoto, X. L. Xia, S. Zhang for collaboration.
This work is supported by NSFC through Grants No. 11535012 and No. 11675041.






\end{document}